\documentclass[11pt]{pazh} 

\usepackage[T2A]{fontenc}
\usepackage{graphics}
\usepackage{graphicx}
\usepackage{latexsym}
\usepackage{amssymb}
\usepackage{amsmath}
\usepackage{hyperref}
\usepackage{soul}

\begin{document}

\title{\bf Evolution of the spiral structure of galaxies from the HST COSMOS field}

\author{V.P. Reshetnikov\inst{1,2}, A.A. Marchuk\inst{1,2}, I.V. Chugunov\inst{1,2},\\
P.A. Usachev\inst{1,2}, and A.V. Mosenkov\inst{3,2}
}

\institute{St.Petersburg State University, Universitetskii pr. 28, St.Petersburg, 
198504 Russia
\and
Pulkovo Astronomical Observatory, Russian Academy of Sciences, St. Petersburg, 
196140 Russia 
\and
Brigham Young University, Provo, USA
}

\abstract{We have investigated the pitch angle ($\psi$) of the spiral arms of galaxies in the 
Hubble Space Telescope COSMOS field. The sample consists of 102 face-on galaxies with 
a two-armed pattern at a mean redshift $\langle z \rangle \approx 0.5$. The typical values of 
$\psi$ in the spiral arms of distant galaxies are shown to be close
to those for nearby spiral galaxies. Within one galaxy the scatter of $\psi$ for different 
arms is, on average, half the mean pitch angle. In the $z$ range from 1 to 0 we have found a 
tendency for $\psi$ to decrease. Our analysis of the $\psi$ distributions in galaxies at 
different redshifts is consistent with the assumption that in most of the
galaxies at $z \leq 0.5$ the spiral arms are tidal in origin or they arose from transient 
recurrent instabilities in their disks.
\keywords{galaxies, photometry, evolution}
}

\authorrunning{Reshetnikov et al.}
\titlerunning{Evolution of the spiral structure}

\maketitle

\section{Introduction}

Spiral arms are the most prominent feature in most of the bright galaxies in the 
surrounding part of the Universe. For example, in the local Universe the
fraction of such galaxies is $\sim$75\% of all the galaxies brighter than an 
absolute magnitude of --20$^m$ in the $B$ passband (Conselice 2006). A huge number of papers are
devoted to investigating the spiral pattern of galaxies (for a recent review, see Sellwood and 
Masters 2022), but many of the questions related to the formation
and maintenance of the existence of spiral arms as well as to their observational 
manifestations remain poorly studied. The importance of studying these
questions stems from the significant influence that the spiral pattern exerts on global 
processes in a galaxy, for example, on the angular momentum redistribution
in the disk (Sellwood and Binney 2002), its stability (Inoue et al. 2021), the star formation 
rate (Querejeta et al. 2021), the chemical evolution (Scarano and L\'epine 2013), etc.

The studies of the spiral pattern of distant galaxies are so far few. It is well known that 
the familiar types of spiral structure in the nearby Universe (grand-design, flocculent, 
multiple arms) were already present at $z \geq 1$ (Elmegreen and Elmegreen 2014). The number 
density of spiral galaxies drops with increasing $z$ (between $z$=0.5 and $z$=2.5
it decreases approximately by an order of magnitude -- Margalef-Bentabol et al. 2022), but 
objects with a grand-design spiral pattern are encountered even at
$z \approx 3$ (Wu et al. 2023).

The main goal of our paper is to investigate the pitch angles in a hundred two-armed spiral 
galaxies up to a redshift $z \sim 1$.

All of the numerical values in the paper are given for the cosmological model with 
$\Omega_m=0.3$, $\Omega_{\Lambda}=0.7$, $H_0 = 70$ km s$^{-1}$ Mpc$^{-1}$.

\section{Sample of galaxies}

To study the structure of distant spiral galaxies, we examined the Hubble Space Telescope (HST)
COSMOS field (Koekemoer et al. 2007). This field with an area of almost 2 deg$^2$ was imaged 
in the F814W filter with ACS. For the selection of objects in COSMOS we used the sample of 
26113 bright (F814W < 22\fm5) galaxies in this field presented in Mandelbaum et al. (2012). 
For all of the objects from the sample we determined the apparent flattening ($b/a$) using 
the SExtractor package (Bertin and Arnouts 1996) and selected 7441 galaxies with 
$b/a \geq 0.7$. Next, we performed a visual examination of the
galaxy images and produced a sample of 184 nearly face-on spiral galaxies with clearly 
distinguishable spiral arms. During our subsequent analysis we left 102 galaxies with a 
distinct two-armed spiral pattern in the final sample. We carried out our further analysis
of the spiral arms of galaxies (see the next section of the paper) based on the galaxy images 
reduced to a scale of 0.$''$03/pixel presented in Mandelbaum et al. (2012).

The sample galaxies were identified with the COSMOS2020 catalog (Weaver et al. 2022). For
each object from COSMOS2020 we took the photometric redshifts (their accuracy is $\sim$1\%) found with
the LePhare code (Ilbert et al. 2006), the absolute magnitudes in the $r$ band (Subaru HSC), 
and the stellar mass estimates in solar masses (M$_{\odot}$). Figures 1a and 1b show the positions 
of the galaxies on the absolute magnitude ($M_r$) -- redshift ($z$) and galactic stellar 
mass -- $z$ planes. It can be seen from the figures that our sample includes galaxies up 
to $z \approx 1$ (the mean redshift of the sample is $\langle z \rangle = 0.47 \pm 0.23$),
with the galaxies being, on average, bright, 
$\langle M_r \rangle = -21\fm85 \pm 0\fm88$, and massive, 
$\langle log\,{\rm M}_* \rangle = 10.52 \pm 0.43$. Observational selection is also clearly seen
in Figs. 1a and 1b: the luminosity and mass of the objects included in the sample increase 
with $z$. If we restrict ourselves only to the most massive galaxies
with $log\,{\rm M}_* \geq 10.5$ (M$_* \geq 3 \times 10^{10}$\,M$_{\odot}$), then such
objects are identified in the entire $z$ range, and the corresponding subsample is relatively 
complete. 

As noted above, galaxies with $b/a \geq 0.7$, which were visually estimated to be 
oriented nearly face-on, were included in the sample. The mean apparent
flattening of the objects in the final sample turned out to be $b/a = 0.87 \pm 0.065$. 
In our further analysis we did not apply the correction for the possible inclination
of the disk plane and assumed the galaxies to be seen exactly face-on.

\begin{figure}
\centering
\includegraphics[width=11.5cm, angle=-90, clip=]{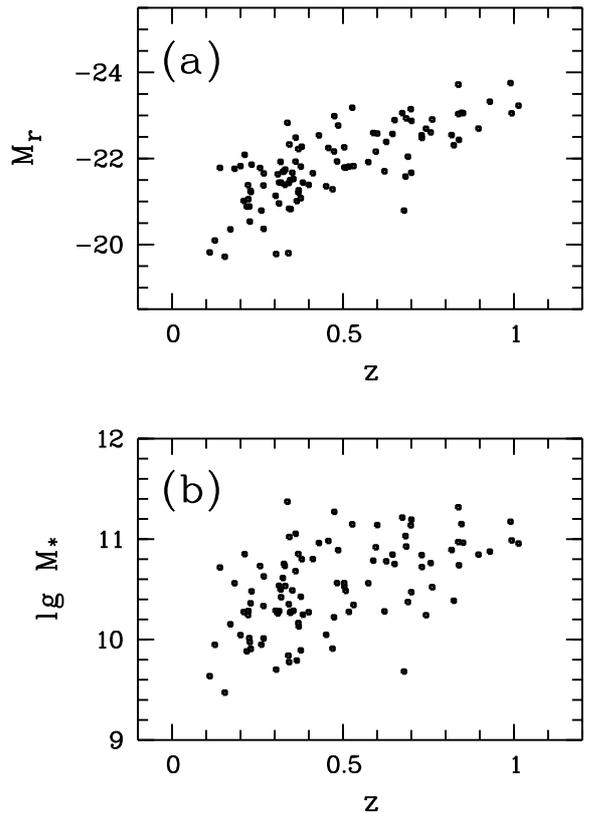}\\
\caption{Redshift dependence of the absolute magnitude of spiral galaxies in the $r$ band (a) 
and their stellar mass (b).}
\end{figure}

\section{Determination of spiral arms parameters}

The procedure of measuring the spiral pattern characteristics is similar to the technique 
from Savchenko et al. (2020) and is briefly described below.
For each galaxy we mark out the individual spiral arms starting from the bar ends or the 
central regions and up to the outer regions, where the spiral is still
visible above the background level. All of the spirals must be clearly distinguishable and 
must not contain any branching and self-intersections. Then, based
on a special algorithm, at each point of the arm we construct a cut perpendicular to the 
direction of the spiral and stretched between the minima of the interarm space. Each cut is 
fitted by an analytically asymmetric Gaussian function in the form
$$I(r) = I_0\times\exp[-\frac{(r-r_{peak})^2}{W_1^2 \cdot s + W_2^2 \cdot (1-s)}],$$ 
where $I_0$ denotes the central flux, $W_{1,2}$ are the inner and outer half-widths, 
$r$ is the distance along the cut, $r_{peak}$ denotes the position of the brightness peak,
and the parameter $s$ is equal to 0 if $r \leq r_{peak}$, otherwise
1 (see Fig. 5 in Savchenko et al. 2020). The resulting profile is convolved with the 
corresponding PSF F814W profile to take into account the influence of
the optical system on the images.

Once the fitting has been performed for all of the cuts, we have a completely measured arm model.
Examples of the resulting marking after our fitting for four galaxies are shown in Fig. 2. 
The positions of the peaks $r_{peak}$ are marked by the dots, while the segment
size corresponds to the derived half-widths.

\begin{figure*}
\centering
\includegraphics[width=12.5cm, angle=0, clip=]{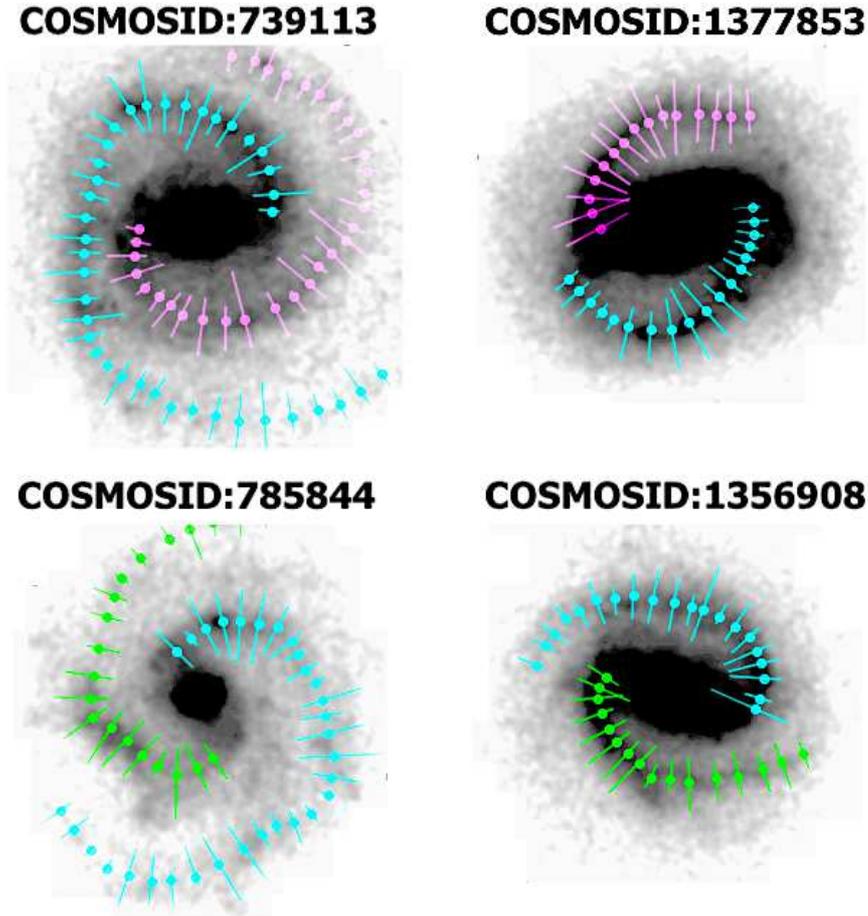}\\
\caption{Examples of galaxies from the sample with measured spirals. The dots mark the positions 
of the brightness peaks; the segments correspond to the spiral half-width. }
\end{figure*}

The pitch angle $\psi$ is estimated based on the constructed cuts across the spiral arm. 
For this purpose, the positions of the peaks along the spiral are plotted
on the $\log r \div \varphi$ plane, where $\varphi$ is the polar angle,
and then the linear regression is found. The slope of the linear dependence is equal to 
the sought-for pitch angle $\psi$, while the root-mean-square (rms) deviation
from the straight line gives an estimate of the error in the pitch angle $\sigma_{\psi}$. 
As a result, in 99 galaxies from the sample we found the pitch angles for both spiral
arms; in three galaxies we managed to perform our measurements only for one of the arms.

\section{Results and discussion}

\subsection{Pitch angles}

Figure 3 shows the distribution of pitch angles averaged over two arms for 99 sample 
galaxies that have measurements for both arms. For comparison, the
distribution for 31 galaxies from three HST deep fields (HDF-N, HDF-S, HUDF) from Savchenko 
and Reshetnikov (2011) is shown in the same figure. The galaxies from Savchenko and 
Reshetnikov (2011) are, on average, at a mean redshift $\langle z \rangle = 0.69 \pm 0.30$,
and their pitch angles found based on a Fourier analysis of images were measured in the F606W
filter. As can be seen from Fig. 3, the distributions for the two samples of distant galaxies, 
in which the pitch angles were determined by different methods, agree well. For example, 
the mean pitch angle for the two-armed spiral galaxies from the COSMOS field
is $\langle \psi \rangle = 14.^{\rm o}85 \pm 5.^{\rm o}51$, while for the galaxies from
Savchenko and Reshetnikov (2011)  $\langle \psi \rangle = 14.^{\rm o}1 \pm 3.^{\rm o}5$.   
If all of the measured arms in COSMOS are considered separately, then the mean value for them
is $\langle \psi \rangle = 15.^{\rm o}02 \pm 7.^{\rm o}05$ (201 spiral arms).

\begin{figure}
\centering
\includegraphics[width=6.5cm, angle=-90, clip=]{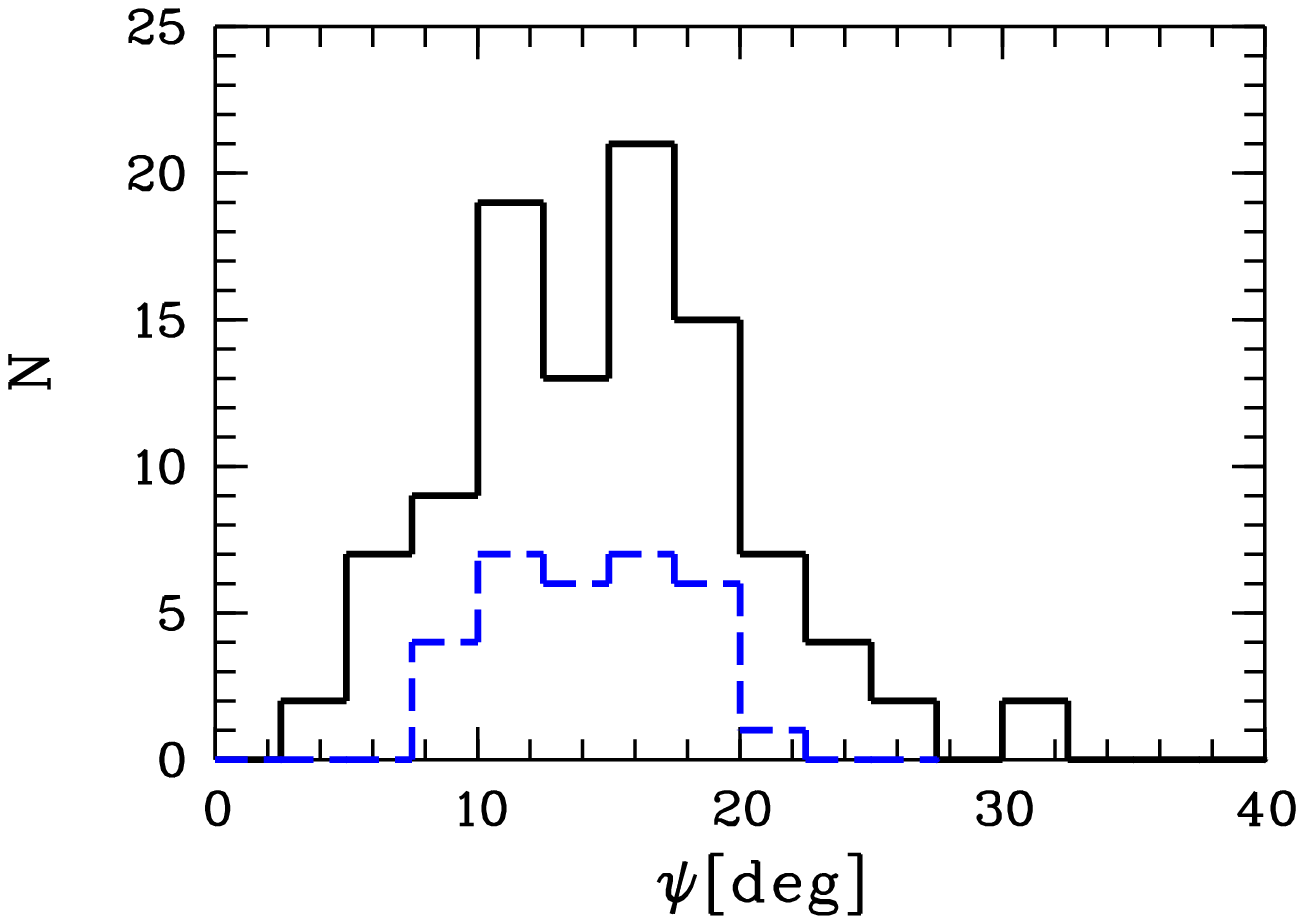}\\
\caption{Distribution of galaxies from the COSMOS field in mean pitch angle (solid line); 
the corresponding distribution for galaxies from three HST deep fields is indicated by the 
blue dashed line (Savchenko and Reshetnikov 2011).
}
\end{figure}

Let us compare the above mean values of $\psi$ with the data for nearby galaxies. 
For example, \\
$\langle \psi \rangle = 15.^{\rm o}2 \pm 3.^{\rm o}7$ (50 grand-design galaxies, the $g$ band;
Savchenko and Reshetnikov 2013), \\
$\langle \psi \rangle = 18.^{\rm o}3 \pm 7.^{\rm o}5$ (79 galaxies, the $r$ band; Yu and 
Ho 2019), \\
$\langle \psi \rangle = 14.^{\rm o}8 \pm 5.^{\rm o}3$  (155 galaxies, the $r$ band; Savchenko
et al. 2020). \\
In 75 nearby grand-design spiral galaxies from Diaz-Garcia et al. (2019) the pitch angle shows
a dependence on morphological type, i.e., it changes from 
$13.^{\rm o}6 \pm 1.^{\rm o}6$ (S0/a--Sab) to $19.^{\rm o}7 \pm 2.^{\rm o}5$ (Scd--Sdm)
(the measurements were made at a wavelength of 3.6 $\mu$m). A similar dependence on 
morphological type is also traceable in other papers (see, e.g., Savchenko et al. 2020). 
In a large sample of objects (4378 galaxies, the r band; Yu and Ho 2020) the pitch
angle varies from $\sim 10^{\rm o}$ for Sa-type galaxies to $\sim 30^{\rm o}$
for Sd (see Fig. 10 in Yu and Ho 2020). 

Thus, the typical pitch angles in galaxies at $z \sim 0.5 - 1$ are close to those for galaxies 
in the surrounding part of the Universe.

\subsection{Scatter of pitch angles}

Within one galaxy the spiral arms shows a fairly large scatter of pitch angles. The mean 
difference in $\psi$ for two spiral arms of the objects from our sample is
$\Delta \psi = | \psi_1 - \psi_2 | = 6.^{\rm o}3 \pm 5.^{\rm o}3$. The mean relative
variation of the pitch angle is $\Delta \psi / \langle \psi \rangle = 
0.48 \pm 0.41$. Thus, the typical scatter of $\psi$ for different arms within
one galaxy reaches half the mean pitch angle. The radial $\psi$ variations within the same 
arm can reach comparable values (Savchenko et al. 2020).

\begin{figure}
\centering
\includegraphics[width=6.5cm, angle=-90, clip=]{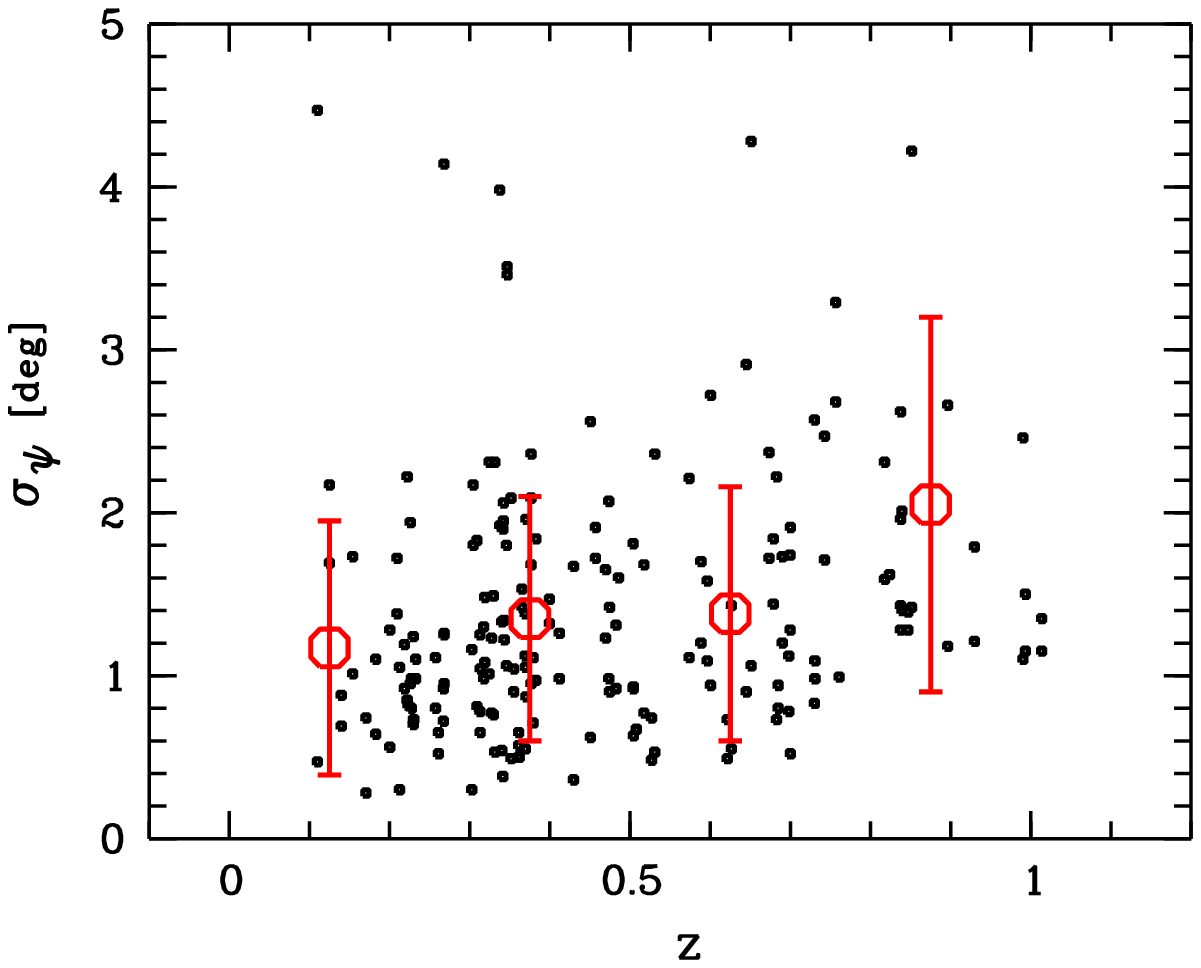}\\
\caption{Redshift dependence of the rms deviation of the spiral arm pitch angle. 
The circles with bars are the mean values in $z$ bins 0-0.25, 0.25-0.5, 0.5-0.75, and 0.75-1.0.
}
\end{figure}

Figure 4 shows the redshift dependence of the error in the pitch angle ($\sigma_{\psi}$) 
for all of the measured spiral arms. The scatter of data points in the figure
is very large, but a certain tendency for $\sigma_{\psi}$ to increase with $z$ is traceable. 
This can imply both the influence of observational selection (the difficulty of
measurements in more distant galaxies) and a greater irregularity, clumpiness of the arms 
in distant galaxies (Elmegreen et al. 2007).

\subsection{Evolution of the pitch angle?}

The change in the mean pitch angles of the spiral arms of galaxies with redshift is shown 
in Figs. 5a and 5b. Despite the large scatter of observational data points, a weak trend of 
$\psi$ with $z$ is noticeable in Fig. 5a. If, alternatively, we consider a subsample
of massive spiral galaxies with $log\,{\rm M}_* \geq 10.5$ less distorted by observational 
selection (Fig. 5b), then this trend becomes more pronounced: from $z$ = 1 to
$z$ = 0 the observed pitch angles decrease, i.e., the spiral arms become, on average, more 
tightly wound. (The statistical correlation between $\psi$ and $z$ in Fig. 5b
is moderate (its Spearman rank correlation coefficient is 0.38), but significant at >99\%.) 
The linear dependence shown in Fig. 5b corresponds to a winding rate $\sim 1^{\rm o}$/Gyr.

\begin{figure}
\centering
\includegraphics[width=12.5cm, angle=-90, clip=]{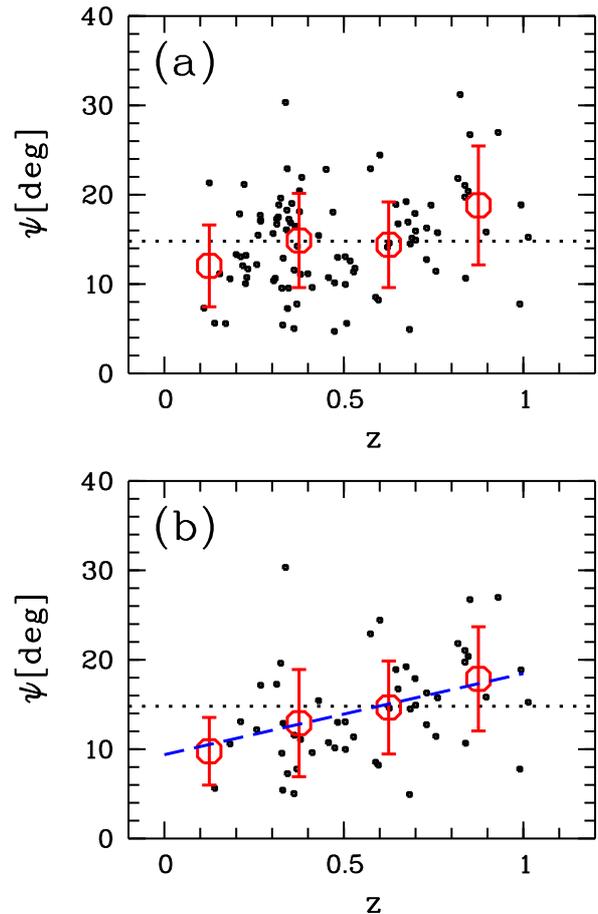}\\
\caption{Redshift dependence of the mean spiral pattern pitch angle: (a) the entire sample 
and (b) the galaxies with $log\,{\rm M}_* \geq 10.5$. The circles with bars in both figures 
are the mean values in $z$ bins 0-0.25, 0.25-0.5, 0.5-0.75, and 0.75-1.0. The
horizontal dotted lines indicate the mean value of $\psi$ for the entire sample, the dashed 
line is the linear regression for massive spiral galaxies.
}
\end{figure}

A possible reason for the existence of the observed trend between $\psi$ and $z$ can be the 
influence of the $k$-correction, i.e., the fact that when passing to more
distant galaxies, we see their images in an increasingly short wavelength range. If the arm 
pitch angle depends on wavelength ($\lambda$) in such a way that $\psi$
increases with decreasing $\lambda$, then this could explain at least in part Fig. 5. 
As said above, our measurements were performed in the HST ACS/F814W filter with
an effective wavelength $\lambda_{eff} \approx 8000$\,\AA, i.e., approximatey
in the $I$ band. When objects at $z = 1$ are observed, this filter will roughly correspond 
to the $B$ band. A direct comparison of the pitch angles for nearby galaxies found in blue 
filters ($B, g$) and in the  near infrared showed no significant differences (see,
e.g., Seigar et al. 2006; Davis et al. 2012; Savchenko et al. 2020). In addition, there is 
evidence that the spiral arm pitch angle can decrease when passing from ``red'' to ``blue'' 
filters (Yu and Ho 2018). If this is the case, then this even enhances the significance of
the observational trend in Fig. 5. Yet another reason can be observational selection when 
selecting objects: among more distant galaxies the probability to miss galaxies with tightly 
wound spiral arms is higher due to the lower spatial resolution. Both these effects,
the influence of the $k$-correction and observational selection, need a further study.
 
It is also worth noting that as the spiral galaxies evolve from $z \sim 1$ to $z \sim 0$, 
the luminosity of the  bulges increases and their contribution to the total luminosity of 
the galaxies grows (see, e.g., Sachdeva et al. 2017). On the other hand, it was noted in
a number of papers that in galaxies with brighter bulges and a higher concentration of the 
luminosity to the center the values of $\psi$ are, on average, lower than those in galaxies 
with fainter bulges and a lower concentration (see, e.g., Savchenko and Reshetnikov 2013; 
Yu and Ho 2020). Thus, the change in the pitch angle with $z$ can at least in part be a
consequence of the evolution of the global structure of galaxies.
 
Different spiral structure generation and maintenance models predict a different behavior of 
the pitch angle as a function of time. For example, in the models in which the arms arise 
from a tidal perturbation and/or transient spiral instabilities in a self-gravitating disk, 
the angle $\psi$ can decrease with time (for a review, see Dobbs and Baba 2014). At the same
time, in the density wave theory (Lin and Shu 1964) the spiral pattern is quasi-stationary 
with an invariable pitch angle.

A simple observational test has recently been proposed for transient and recurrent spiral 
arms (Pringle and Dobbs 2019). Let the spiral arm during its formation have a pitch angle 
$\psi_{max}$, the arm then gradually winds up and, finally, disappears at some minimum
$\psi_{min}$. Then, based on simple reasoning, we can find that the cotangent of the pitch 
angle changes linearly with time: $ctg\,\psi \propto t$. Consequently, if at an arbitrary
instant of time we will consider a random sample of galaxies in which the spiral arms are 
at different winding stages, then one might expect the galaxies to be distributed uniformly 
in $ctg\,\psi$ in the range from $ctg\,\psi_{min}$ to $ctg\,\psi_{max}$. The application of 
this test for two samples of nearby galaxies, to a first approximation, confirmed the 
uniformity of the $ctg\,\psi$ distribution (Pringle and Dobbs 2019; Lingard et al. 2021).

\begin{figure*}
\centering
\includegraphics[width=11.5cm, angle=-90, clip=]{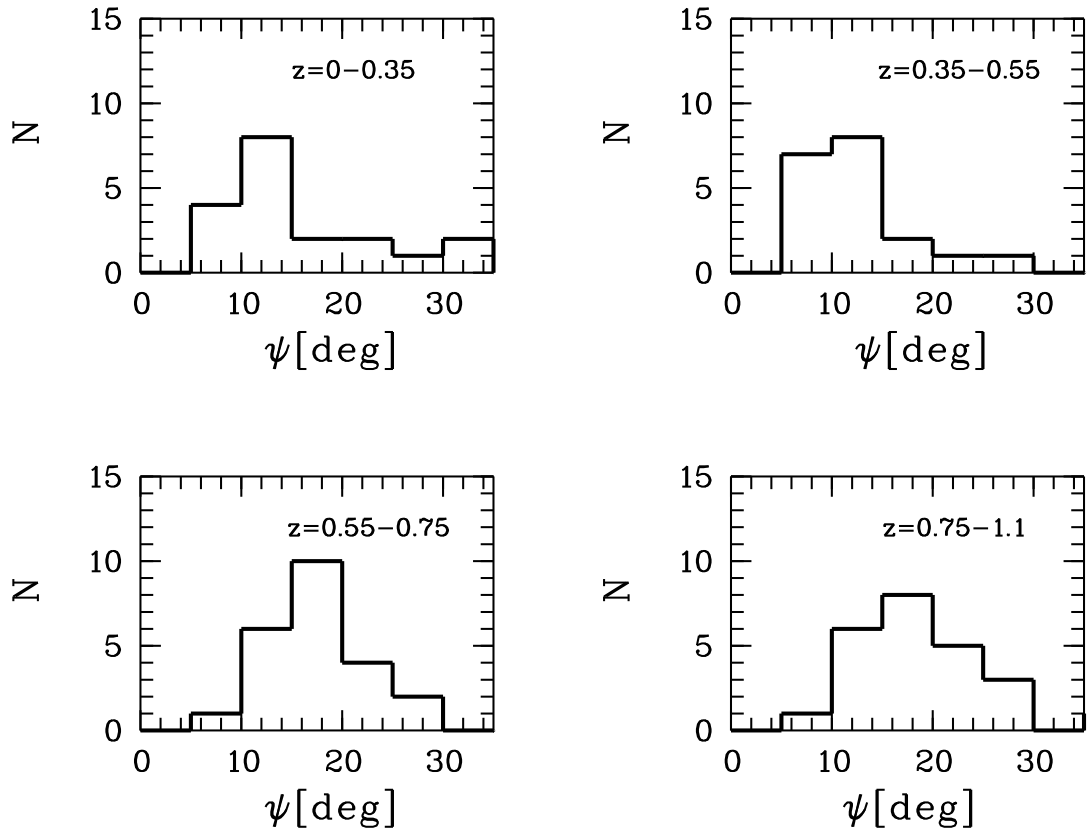}\\
\caption{Distributions of spiral arm pitch angles $\psi$ for the massive galaxies from the 
sample ($log$\,M$_* \geq 10.5$) in different redshift bins.
}
\end{figure*}

Let us consider how the distant galaxies in the COSMOS field are distributed in $ctg\,\psi$. 
We took a relatively complete subsample of galaxies with $log$\,M$_* \geq 10.5$, for which 
there are $\psi$ measurements for 85 individual spiral arms. Figures 6 and 7 present
the distributions of galaxies in $\psi$ and $ctg\,\psi$ in four $z$ bins. (The bins were 
chosen in such a way that approximately the same number of galaxies fell into
them.) To a first approximation, the distributions shown in Fig. 6 are nonuniform, and 
their means are shifted toward smaller $\psi$ with decreasing redshift. When passing to 
$ctg\,\psi$, the shape of the distributions changes (Fig. 7): in the first two redshift bins the
distributions are nearly uniform (of course, within the limits of poor statistics), in agreement 
with the results of recent works in which nearby galaxies were studied (Pringle and Dobbs 2019; 
Lingard et al. 2021). At $z > 0.5$ the distributions exhibit global peaks at
$ctg\,\psi \approx 3$ ($\psi \approx 18^{\rm o}$). The change in the shape of
the distributions with redshift can imply that different spiral pattern generation mechanisms 
could prevail at different epochs.

\begin{figure*}
\centering
\includegraphics[width=11.5cm, angle=-90, clip=]{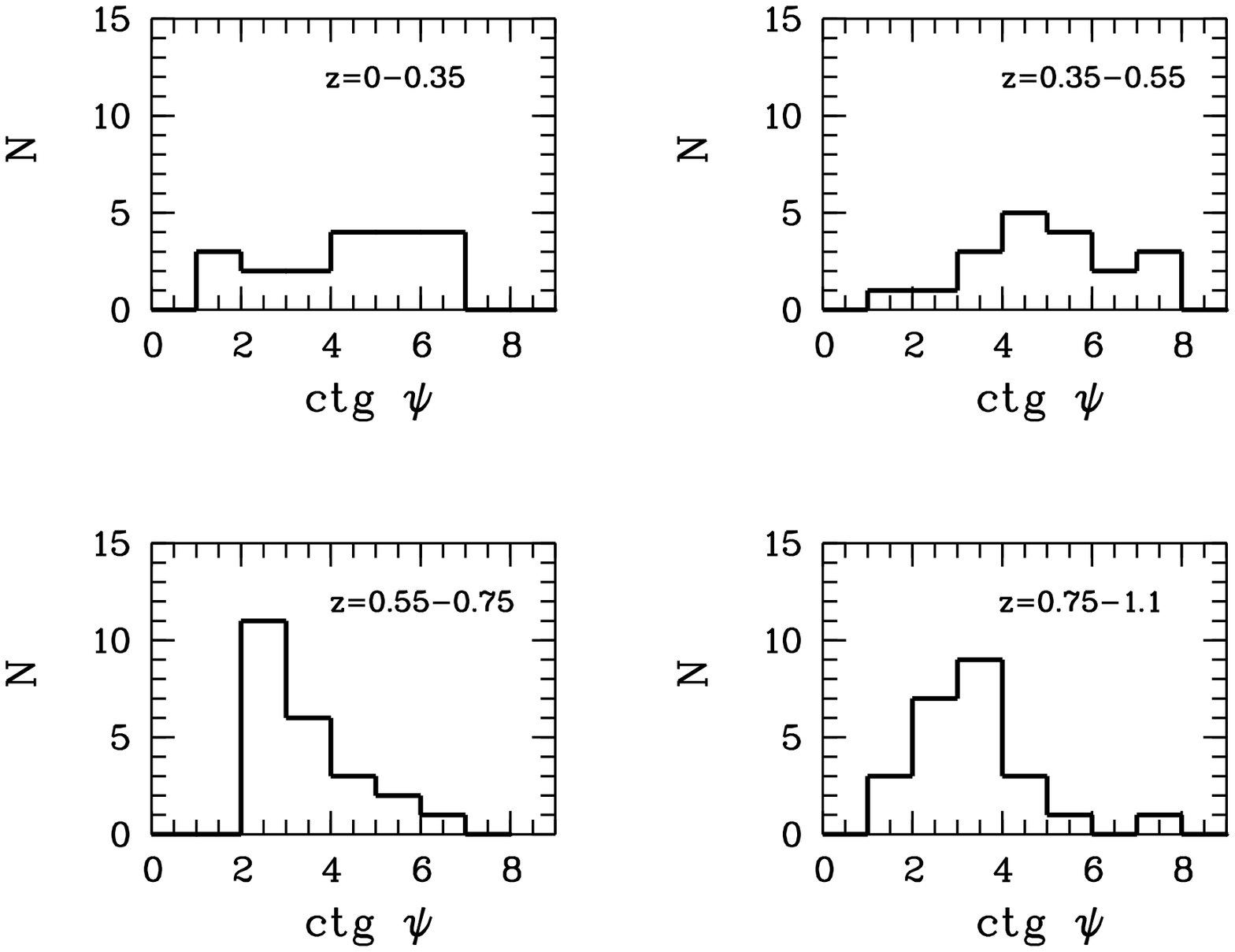}\\
\caption{Distributions of $ctg\,\psi$ for the spiral arms of the massive galaxies from 
the sample ($log$\,M$_* \geq 10.5$) in different redshift bins.
}
\end{figure*}

\section{Conclusions}

In our paper for the first time we have considered in detail the shape of the spiral 
arms in two-armed galaxies in the HST COSMOS field.

We found that the typical values of the spiral pattern pitch angle up to $z \sim 1$, 
$\langle \psi \rangle \approx 15^{\rm o}$, are close to those for nearby galaxies.

Within one galaxy different spiral arms exhibit greatly differing pitch angles. For 
individual arms the error in $\psi$ increases with $z$, which may be a consequence of their 
growing irregularity.

We found an observational trend suggesting a gradual decrease in $\psi$ with decreasing 
$z$ (Fig. 5). This trend corresponds to a mean winding rate $\sim 1^{\rm o}$/Gyr.

Our analysis of the $ctg\,\psi$ distributions (the Pringle-Dobbs test) in massive galaxies 
($log$\,M$_* \geq 10.5$) for different redshift bins is consistent with the fact that
at $z \leq 0.5$ tidal perturbations and transient instabilities in their disks could be 
the main spiral arm generation mechanisms. Applying this test for large
samples of spiral galaxies at different redshifts can become a useful tool for studying 
the evolution of the spiral pattern.

Note that our results are based on a relatively small sample of objects, and they need to be 
confirmed on much more extensive material. A combination of the data from HST and JWST deep 
fields with the development of computer image analysis methods will soon make it possible to 
investigate the questions touched on in our paper in much more detail.

\section{Funding}

This work was supported by the Russian Science Foundation (project no. 22-22-00483). \\

\begin{center}
{\Large \bf References}
\end{center}

\small

\noindent
E. Bertin and S. Arnouts, Astron. Astrophys. Suppl.
Ser. 117, 393 (1996).\\

\noindent
C.J. Conselice, Mon. Not. R. Astron. Soc. 373, 1389
(2006).\\

\noindent
B.L. Davis, J.C. Berrier, D.W. Shields, et al., Astrophys. J. Suppl. Ser. 199, 33 (2012).\\

\noindent
S. Diaz-Garcia, H. Salo, J. H. Knapen, and M. Herrera-Endoqui, Astron. Astrophys. 631, A94
(2019).\\

\noindent
C. Dobbs and J. Baba, Publ. Astron. Soc. Austr. 31, e035 (2014).\\

\noindent
D.M. Elmegreen, B.G. Elmegreen, T. Ferguson, and B. Mullan, Astrophys. J. 663, 734 (2007).\\

\noindent
D.M. Elmegreen and B.G. Elmegreen, Astrophys. J. 781, 11 (2014).\\

\noindent
O. Ilbert, S. Arnouts, H. J. McCraken, et al., Astron. Astrophys. 457, 841 (2006). \\

\noindent
S. Inoue, T. Takagi, A. Miyazaki, et al., Mon. Not. R. Astron. Soc. 506, 84 (2021).\\

\noindent
A.M. Koekemoer, H. Aussel, D. Calzetti, et al., Astrophys. J. Suppl. Ser. 172, 196 (2007). \\

\noindent
C.C. Lin and F.H. Shu, Astrophys. J. 140, 646 (1964). \\

\noindent
T. Lingard, K.L. Masters, C. Krawczyk, et al., Mon. Not. R. Astron. Soc. 504, 3364 (2021). \\

\noindent
R. Mandelbaum, Ch.M. Hirata, A. Leauthaud, R.J. Massey, and J. Rhodes, Mon. Not. R. Astron.
Soc. 420, 1518 (2012).\\

\noindent
B. Margalef-Bentabol, Ch.J. Conselice, B. Haeussler, et al., Mon. Not. R. Astron. Soc. 511, 1502
(2022). \\

\noindent
J.E. Pringle and C.L. Dobbs, Mon. Not. R. Astron. Soc. 490, 1470 (2019).\\

\noindent
M. Querejeta, E. Schinnerer, S. Meidt, J. Sun, A.K. Leroy, et al., Astron. Astrophys. 656, A133
(2021). \\

\noindent
S. Sachdeva, K. Saha, and H.P. Singh, Astrophys. J. 840, 79 (2017). \\

\noindent
S.S. Savchenko and V.P. Reshetnikov, Astron. Lett. 37, 817 (2011). \\

\noindent
S.S. Savchenko and V.P. Reshetnikov, Mon. Not. R. Astron. Soc. 436, 1074 (2013). \\

\noindent
S. Savchenko, A. Marchuk, A. Mosenkov, and K. Grishunin, Mon. Not. R. Astron. Soc. 493, 390 
(2020). \\

\noindent
S. Scarano and J. R. D. L\'epine,  Mon. Not. R. Astron. Soc. 428, 625 (2013).\\

\noindent
M.S. Seigar, J.S. Bullock, A.J. Barth, and L.C. Ho, Astrophys. J. 645, 1012 (2006).\\

\noindent
J.A. Sellwood and K.L. Masters, Ann. Rev. Astron. Astrophys. 60, 73 (2022).\\

\noindent
J.A. Sellwood and J.J. Binney, Mon. Not. R. Astron. Soc. Astrophys. 336, 785 (2002).\\

\noindent
J.R. Weaver, O.B. Kauffmann, O. Ilbert, et al., Astrophys. J. Suppl. Ser. 258, 11 (2022).\\

\noindent
Y. Wu, Z. Cai, F. Sun, et al., Astrophys. J. Lett. 942, id.L1 (2023).\\

\noindent
S.-Yu. Yu and L.C. Ho, Astrophys. J. 869, 29 (2018).\\

\noindent
S.-Yu. Yu and L.C. Ho, Astrophys. J. 871, 194 (2019).\\

\noindent
S.-Yu. Yu and L.C. Ho, Astrophys. J. 900, 150 (2020).

\end{document}